\documentclass[twocolumn,10pt,journal]{IEEEtran}

\title{Increasing Smart Meter Privacy Through\\ Energy Harvesting and Storage Devices}

\author{\IEEEauthorblockN{Onur Tan\IEEEauthorrefmark{1}, Deniz G\"{u}nd\"{u}z\IEEEauthorrefmark{2}, H. Vincent Poor,\IEEEauthorrefmark{3}}\\

\IEEEauthorblockA{\IEEEauthorrefmark{1}Centre Tecnol\`ogic de Telecomunicacions de Catalunya (CTTC), Barcelona, Spain.}\\
\IEEEauthorblockA{\IEEEauthorrefmark{2}Department of Electrical and Electronic Engineering, Imperial College London, London, UK.}\\
\IEEEauthorblockA{\IEEEauthorrefmark{3}Department of Electrical Engineering, Princeton University, Princeton, NJ, USA.}

\thanks{This work was supported in part by the Spanish Government under project TEC2010-17816 (JUNTOS), and in part by the U.S. National Science Foundation under Grant CCF-1016671. This work was presented in part at the 2012 IEEE International Conference on Smart
Grid Communications in the Cognitive and M2M Communications and Networking
for Smart Grid Workshop.}
}

\usepackage{amsfonts}
\usepackage{amssymb}
\usepackage{amsmath}
\usepackage{dsfont}
\usepackage{amscd}
\usepackage{graphicx, subfigure}
\usepackage{psfrag}
\usepackage{xcolor}

\usepackage{cite}
\usepackage{epsfig}

\begin{document}

\maketitle
\thispagestyle{empty}
\pagestyle{empty}

\begin{abstract}
Smart meters are key elements for the operation of smart grids. By providing near realtime information on the energy consumption of individual users, smart meters increase the efficiency in generation, distribution and storage of energy in a smart grid. The ability of the utility provider to track users' energy consumption inevitably leads to important threats to privacy. In this paper, privacy in a smart metering system is studied from an information theoretic perspective in the presence of energy harvesting and storage units. It is shown that energy harvesting provides increased privacy by diversifying the energy source, while a storage device can be used to increase both the energy efficiency and the privacy of the user. For given input load and energy harvesting rates, it is shown that there exists a trade-off between the information leakage rate, which is used to measure the privacy of the user, and the wasted energy rate, which is a measure of the energy-efficiency. The impact of the energy harvesting rate and the size of the storage device on this trade-off is also studied.
\end{abstract}


\section{Introduction}
\label{sec:intro}
A smart grid (SG) is an energy network that manages and controls energy generation and distribution more efficiently and intelligently by following the users' energy demands in real-time through computer and communication technologies. Transition from traditional power grids to SGs are expected to have a revolutionary effect on future energy networks~\cite{GridOfFuture},~\cite{PrivacyChallenge}. SGs can yield energy efficiency through savings in generation and transmission of energy, reduce costs on both the user and the utility provider (UP) sides, and increase reliability and robustness. They also provide important environmental benefits by reducing the carbon footprint and integrating renewable energy sources into the energy network. Introducing alternative energy sources and energy storage devices into the network will significantly reduce the load on the energy network and improve its efficiency. For instance, plug-in electric vehicles on the distribution grid can be used for distributed energy storage by means of their rechargeable batteries (RBs)~\cite{GridOfFuture}. Similarly, renewable energy sources can be integrated into the energy network through energy harvesting (EH) devices, which can generate energy from ambient sources such as solar, thermal or wind, and reduce the users' dependence on the grid~\cite{PowerManagementinEH}.

To exploit these potential benefits, the components of an SG are connected through a two-way communication network that allows the exchange of information in real time among the users and the UP. This enables real-time optimization of load management in SGs~\cite{SmartGridCom}. An important component of this critical data network for SGs is the advanced metering system. Smart meters (SMs) are communication devices that measure the energy consumption of the users and transmit their readings to the UP in real time. Currently, a typical smart-meter reports the energy consumption readings to the UP every $15$ minutes; however, the measuring frequency is expected to increase in the near future to provide near real-time energy consumption data to the UP. Significant energy savings have been reported even solely based on the user's increased awareness of his/her real-time energy consumption~\cite{DynamicEnergyConsumption}. However, despite their potential for increasing the efficiency of energy distribution networks, SG technologies, in particular smart metering systems, raise important privacy and security concerns for the users~\cite{PrivacyChallenge},~\cite{NewEnergy},~\cite{CoordinatedDataInjection}.

\begin{figure}
\hspace{-0.3cm}
\includegraphics[scale=0.31]{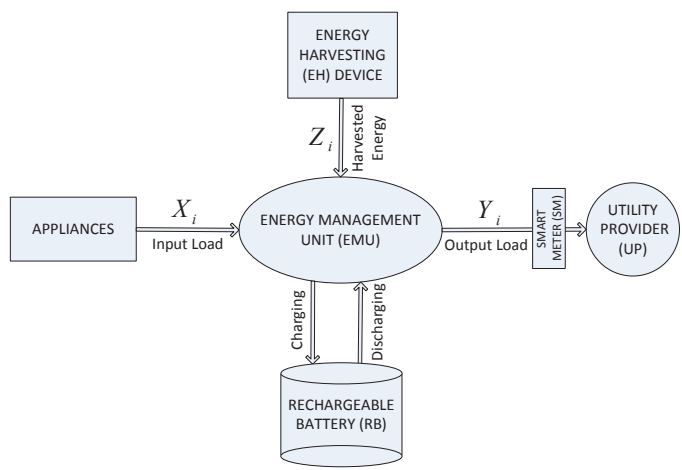}
\caption{A smart-meter (SM) system diagram with energy and information flows. The user, in addition to its connection to the energy grid, also has an EH device and an RB at its use. The energy flow in the system is managed by the energy management unit (EMU). The SM reads only the energy that is supplied by the UP at each interval. The readings are reported to the UP correctly without any tempering, but potentially in an encrypted manner.}
\label{fig:GeneralSystem}
\end{figure}

SM data can be easily analyzed for surveillance purposes by tracking appliance usage patterns, employing nonintrusive appliance load monitors and data mining algorithms~\cite{Nonintrusive},~\cite{Taxonomy},~\cite{DataMining}. At the very least, through SM readings it is possible to infer whether a user is at home or not. But, through more advanced pattern recognition techniques, energy consumption patterns of individual appliances can be identified with high accuracy even when the SM can read only the aggregated household energy consumption~\cite{NeuronBased}. As a striking example,~\cite{MultimediaContent} illustrates the possibility of detecting the channel displayed on a television, and even identifying the content, just by analyzing the power profile of the household. Even assuming that the SM readings are transmitted to the UP in an encrypted manner, preventing third parties from accessing the user's private energy consumption data, the UP will receive significant personal information about the user. Thus, even if only partially, assuring the privacy of the household's electrical load profile is essential for users.

In this work, we study SM privacy from the fundamental information theoretic perspective. We measure the privacy of the user's energy profile with respect to the UP in terms of the $\emph{information leakage rate}$, which denotes the mutual information rate between the real energy consumption of the appliances and the SM readings. Using Shannon entropy to measure privacy is not new. Minimizing the information leakage rate is equivalent to maximizing the $\emph{equivocation}$, which was introduced by Shannon in~\cite{SecrecyShannon} in the context of secure communications. Mutual information has previously been proposed as a measure of privacy in SMs in~\cite{ElecPrivacy},~\cite{UtilityPrivacyData},~\cite{RechargeableBattery} and~\cite{CombatingUnauthorizedLoadSignal}. Modeling the input load as a discrete time random process, information leakage rate measures the amount of information the UP learns about the input load after observing the output load, i.e., the energy requested by the user. We assume that the UP may know the statistics of the input load as well as the stochastic behavior of the energy management policy; however, it cannot observe the input load or harvested energy directly. The UP has to estimate the realization of the input load based on its statistical knowledge and its observation of the output load. The user wants to minimize the information leakage rate to achieve the highest level of privacy. While cryptographic algorithms rely on mathematical operations and the complexity of their computation by using encryption keys, information theoretic security does not depend on encryption keys and assures reliable privacy regardless of the computational power of an intruder, the UP in our case~\cite{InformationTheoreticSecurity}.

Building on our previous work~\cite{SMPrivacyOurConferencePaper}, we study the privacy of an SM system from the perspective of a single user. In our system model, depicted in Fig.~\ref{fig:GeneralSystem}, we integrate an EH device as an alternative energy source and an RB as an energy storage unit. The energy flow is managed by the energy management unit (EMU). We consider a discrete time system. At each time instant $i$, the appliances request a certain amount of energy, denoted by $X_{i}$. This amount is reported to the EMU which is responsible for providing this exact amount to the appliances; that is, we do not allow energy outages or rescheduling of appliance operations in this work. We also consider only the real power consumption of the devices and assume that the SM only reads and reports this quantity. Moreover, we also ignore inefficiencies and mismatches in providing the energy requirement of the appliances from different energy sources, and consider only the energy that is consumed by the appliances. The EMU has access to three different energy sources$\colon$the energy grid, the EH device and the energy storage unit. At any time instant it can provide the energy requested by the appliances from one or more of these sources. The goal of the EMU is to increase both the energy efficiency of the system and the privacy of the user.

We employ stochastic battery policies based on the harvested energy, energy demand of the appliances and the state of the storage unit. We model the energy generation profile of an EH device as a stochastic process whose behavior depends on the characteristics of the underlying energy source and the device itself. Therefore, it is likely that the harvested energy sometimes does not match the energy required by the system and the extra energy would be wasted if not stored. Introducing an RB for energy storage into the system is essential for better utilization of the harvested energy. On the other hand, considering the increasing use of alternative energy sources (such as solar panels) by households, and the availability of rechargeable storage units (such as electric vehicles) with significantly large storage capacities, it is meaningful to exploit these devices not only to decrease the dependency on the SG and to increase the energy efficiency, but also to provide additional privacy for the users. The equivocation of the UP about the real energy consumption can be manipulated by charging and discharging the RB and by using the harvested energy. Hence, the benefits of the RB are twofold$\colon$$i)$ it can increase the energy efficiency of the system by storing extra harvested energy; and $ii)$ it can increase the privacy of the user by hiding the energy consumption profile from the UP. We show in this paper that there exists a trade-off between energy efficiency and privacy for the optimal EMU operation, and the operating point on this trade-off can be chosen based on the privacy sensitivity of the underlying input load and the cost of energy.

The main contributions of this work can be summarized as follows$\colon$
\begin{enumerate}

\item We introduce an energy efficiency-privacy trade-off in a smart meter system considering the availability of an EH device and an RB. To the best of our knowledge, this is the first work that provides an analytical study on the effect of an alternative energy source on SM privacy.

\item Focusing on a discrete-time system model we study the effect of energy harvesting rate on the energy efficiency-privacy trade-off.

\item We illustrate numerically that the increased battery capacity significantly reduces the information leakage rate.

\item While no grid energy is allowed to be wasted in the above analysis, we also study the increased privacy that can be achieved by wasting the grid energy for very sensitive applications.

\end{enumerate}

We use the following notation in the rest of the paper. Random variables are denoted with uppercase letters, e.g., $X$, and their realizations are denoted with lowercase letters, e.g., $x$. A random variable takes values from a finite set $\mathcal{X}$ following a probability mass function $p_{X}(x)$. The subscript $X$ will be omitted when it is obvious from the context. An n-length random sequence is denoted by $X^n=X_{1},\ldots,X_{n}$. $E[X]$ denotes the expectation of the random variable $X$. The entropy of a random variable $X$ is defined by
\begin{align}
\label{eq:entropy}
H(X) \triangleq - \sum_{x \in \mathcal{X}} p(x) \log p(x).
\end{align}

\noindent $H(\cdot|\cdot)$ and $H(\cdot,\cdot)$ denote conditional entropy and joint entropy, respectively, which are defined similarly. The mutual information between random variables $X$ and $Y$ is defined as
\begin{align}
\label{eq:mutualinf}
I(X;Y) = H(X) - H(X|Y).
\end{align}

The rest of the paper is organized as follows. In Section~\ref{s:Related Work}, we summarize some of the related work on privacy issues in SM systems. In Section~\ref{s:SystemModel}, we introduce the system model. Section~\ref{s:InformationLeakageRatesec} describes the technique to compute the information leakage rate. In Section~\ref{s:ResultsandObservations}, we present our results and compare them with the existing results in the literature. Finally, we conclude our work in Section~\ref{s:ConclusionsFutureWork}.

\section{Related Work}\label{s:Related Work}
In recent years SMs have gained increasing popularity with growing support from the UPs and governments with the promise of increased energy efficiency. This also has raised privacy issues, and the literature in this field is growing rapidly. Various techniques have recently been proposed to provide a certain level of privacy for SM users. Anonymization~\cite{Anonymization}, aggregation~\cite{Aggregation}, homomorphism~\cite{Homomorphism} and obfuscation~\cite{Obfuscation} are some of the techniques that have been studied in the literature. In~\cite{DifferentialPrivacy}, the authors present a method for establishing privacy assurances in terms of differential privacy, i.e., RB is used to modify the energy consumption by adding or subtracting noise and thereby, the energy consumption of the individual appliances can be hidden. Moreover, they also consider various constraints on the RB such as capacity and throughput. In~\cite{ProtectingConsumerPrivacy} a method to provide privacy against potential non-intrusive load monitoring techniques is proposed. A non-intrusive load-leveling algorithm is used to flatten the consumption of the user by means of an RB. Similarly,~\cite{CombatingUnauthorizedLoadSignal} proposes three techniques, i.e., fuzzing, targeted entropy maximization and targeted fuzzing. The authors intend to obfuscate the load by masking the individual loads with the use of an RB. Basically, fuzzing changes the load randomly over an interval, the targeted entropy maximization technique chooses the desired load level that maximizes the entropy of possible individual events, and targeted fuzzing builds a probability distribution to do so.

Most of the earlier work on SM privacy assumes that the user has control over the smart-meter readings and can manipulate these readings before sending the data to the UP. For example, Bohli et al.~\cite{Aggregation} propose sending the aggregated energy consumption of a group of users to the UP. Li et al.~\cite{Li:SmartGridComm:10} consider using compressed sensing techniques for the transmission of the SM reading of active users based on the assumption that SM data transmission is bursty. Bartoli et al.~\cite{SecureLosslessAggregation} propose data aggregation together with encryption to forward smart meter readings. Marmol et al.~\cite{AdditiveHomomorphicEncryption} propose using \textquotedblleft additively homomorphic encryption\textquotedblright, which allows the UP to decode only the total energy consumption of a group of users while keeping the individual readings secure. Rajagopalan et al.~\cite{Sankar} propose compression of the smart-meter data before being transmitted to the UP. Unlike this line of research, we assume that the SM reads the amount of energy that the user gets from the grid at each time interval and the meter readings are reported to the UP without being tempered by the user. Hence, privacy in our model is achieved by differentiating the output load, i.e., the energy received from the UP, from the input load, i.e., the real energy consumption of the user, as much as possible.

A similar approach has been taken in some other previous work as well. RBs have been proposed to partially obscure the energy consumption of the user in~\cite{ElecPrivacy},~\cite{RechargeableBattery},~\cite{DifferentialPrivacy},~\cite{ProtectingConsumerPrivacy} and~\cite{ApplianceLoad}. The main goal of the proposed energy management algorithms in these papers is to protect the privacy of the user. References~\cite{ElecPrivacy} and~\cite{ApplianceLoad} study variational distance, cluster similarity and regression analysis to measure privacy and propose various heuristic techniques, such as the best-effort and power mixing algorithms. A discrete-time system model is considered in~\cite{RechargeableBattery} and stochastic battery policies are studied with mutual information between the input and output loads as the measure of privacy. In~\cite{PrivacyAlternativeEnergySource} a similar information theoretic privacy analysis is carried out in the presence of an EH device that can provide energy limited by peak and average power constraints.

\section{System Model}\label{s:SystemModel}
We study the energy input/output system illustrated in Fig.~\ref{fig:GeneralSystem} under a discrete-time system model. The input load $X_{i}$ represents the total energy demand of the appliances at time instant $i$. The output load $Y_{i}$ denotes the amount of energy that the system requests from the UP, while $Z_{i}$ denotes the amount of harvested energy at time instant $i$. We assume that there is a minimum unit of energy; and hence, at each time instant $i$, the input load, harvested energy and output load are all integer multiples of this energy unit. Over time, we assume that the input load $X^n= X_1, X_2,\ldots,X_n$ is an independent and identically distributed (i.i.d.) sequence with marginal distribution $p_{X}$ over $\mathcal{X}=\{0,1,\ldots,N\}$. The harvested energy is also modelled as a discrete time stochastic process, where $Z^n = Z_1, Z_2, \ldots, Z_n$ is an i.i.d. sequence with marginal distribution $p_{Z}$ over $\mathcal{Z}=\{0,1,\ldots,M\}$. The characteristics of the EH distribution, $p_{Z}$, depend on the design of the energy harvester. For example, for a solar energy harvester the average harvested energy can be increased by scaling the size and the efficiency of the solar panel. Note that the energy consumed by the appliances and the harvested energy are independent of each other.

The output load is the amount of energy that is demanded from the UP, and is denoted by $Y^n = Y_1, Y_2, \ldots, Y_n$ with $Y_{i}$ taking values in $\mathcal{Y}=\{0,1,\ldots,L\}$. We denote the energy in the battery at time instant $i$ by $B_i$. We assume that the RB has a maximum capacity of $K$ energy units, i.e., $B_i \leq K$, $\forall i$, while the system is not bounded by the maximum amount of energy that can be provided by the UP, i.e., $L \geq (N+K)$$\footnote{The energy we consider in this model is the real energy measured by the smart meter and we ignore the reactive power or the power factor which can also be used to make deductions about the input load. Moreover, we also assume that the energy demand of the appliances is satisfied by transferring an equivalent amount of energy from the RB, EH unit or UP; that is, we do not consider the effect of the supply voltage, frequency or the characteristics of the appliances on the amount of energy that needs to be requested from the corresponding energy source. Such quantities could also be incorporated into our model by considering vector-valued measurements, but this added complexity is not necessary for studying the fundamental trade-offs considered here.}$.

We consider stochastic energy management policies at the EMU that depend on the instantaneous input load, harvested energy and the battery state. An energy management policy maps the energy requested by the appliances, $X_i$, the harvested energy, $Z_i$, and the battery state, $B_{i-1}$, to the output load, $Y_i$, and the next battery state, $B_i$. Note that in general a larger set of energy management policies is possible. The EMU can decide its actions based on all the past input/output loads, harvested energy amounts and the battery states. For example~\cite{RechargeableBattery} considers policies that take into account the previous output load, $Y_{i-1}$. Similarly, the best effort policy proposed in~\cite{ApplianceLoad}, in which the EMU aims to keep the output load value as stable as possible, is simply a special case of the battery/output load conditioned policies in~\cite{RechargeableBattery}. To keep the complexity of possible energy management policies simple, we restrict our attention to energy management policies that depend only on $(X_i,Z_i,B_{i-1})$, and satisfy
\begin{align}
\label{eq:inputload}
 Z_{i}+(B_{i}-B_{i-1})+Y_i\geq X_i,
\end{align}
\noindent which guarantees that the energy demand of the appliances is always satisfied.

We assume that the SM provides the output load $Y_{i}$ at each time instant to the UP perfectly. That is, we do not allow the user to manipulate the SM reading. Moreover, we also assume that $p_{X}$ and $p_{Z}$ are known by the UP, whereas no information about the realizations of either the input process $x^{n}$, or the EH process $z^{n}$, is available at the UP, which observes only the output load, $y^{n}$. The equivocation, $H(X^n|Y^n)$, measures the uncertainty of the UP about the real energy consumption after observing the output load. We have,
\vspace{0.2cm}\noindent
\begin{align}
\label{eq:mutualinfo}
H(X^n|Y^n)&=H(X^n)-I(X^n;Y^n).
\end{align}

\vspace{0.2cm}\noindent Since $H(X^n)$ is a characteristic of the appliances and is assumed to be known, the EMU tries to minimize $I(X^n;Y^n)$ in order to maximize the equivocation. Accordingly, the privacy achieved by an energy management policy is measured by the \emph{information leakage rate}, defined as
\vspace{0.1cm}\noindent
\begin{align}
\label{eq:leakage}
I_{p}&\triangleq\lim_{n\to\infty}\frac{1}{n}I(X^n;Y^n),
\end{align}

\vspace{0.2cm}\noindent where $X^n=(X_{1},X_{2},\ldots,X_{n})$, $Y^n=(Y_{1},Y_{2},\ldots,Y_{n})$, and $I(X^n;Y^n)$ is the mutual information between vectors $X^n$ and $Y^n$.

Due to the finite capacity of the RB and the stochastic nature of the input and EH processes, some of the harvested energy will be wasted. To measure the proportion of the energy wasted by an energy management policy, we define the \emph{wasted energy rate} as follows:
\vspace{0.1cm}\noindent
\begin{align}
\label{eq:wastedenergyrate}
E_{w}\triangleq\lim_{n\to\infty}\frac{1}{n}\sum\limits_{i=1}^{n}{(Z_{i}+Y_{i}-X_{i})}.
\end{align}

\vspace{0.2cm}We say that an information leakage-wasted energy rate pair $(I_p, E_w)$ is \textit{achievable} if there exists an energy management policy satisfying (\ref{eq:leakage}) and (\ref{eq:wastedenergyrate}). The closure of the set of all achievable rate pairs is called the \textit{rate region} $\Gamma$. In general the energy management policy that minimizes the information leakage rate does not necessarily minimize the wasted energy rate. From the classical time-sharing arguments \cite{ElementsInformationTheory} we can readily see that the rate region $\Gamma$ is convex. Since the region is also closed by definition, it is sufficient to identify the boundary of region $\Gamma$, which characterizes the optimal trade-off between privacy and energy efficiency.

To illustrate the privacy benefits of having an EH device, we first consider a system without an RB. In this case, the EMU uses as much as possible from the harvested energy, and asks for energy from the UP only when the harvested energy is not sufficient. Therefore, we can define $Y_{i}$ as a deterministic function of $X_{i}$ and $Z_{i}$ as follows:
\vspace{0.2cm}
\begin{align}
\label{eq:outputload}
  Y_{i} =(X_{i}-Z_{i})^+ \triangleq \left\{ \begin{array}{ll}
  X_{i}-Z_{i},&\mbox{if $X_{i}-Z_{i}>0$},\\
  0, & \mbox{if $X_{i}-Z_{i}\leq0$}.\\
  \end{array}\right.\\ \nonumber
\end{align}

In general, it is possible to ask for energy from the UP even when $X_{i}=0$. This will increase the privacy by confusing the UP, but waste energy. We do not allow wasting energy from the UP unless otherwise stated, as this would be costly in practical systems. Obviously, when there is no harvested energy, i.e., $\mathrm{Pr}\{Z=0\}=1$, then we have $Y_{i}=X_{i}$ for $\forall i$, and $I_{p}=\frac{1}{n}H(X^n)=H(X)$, i.e., the UP knows the input load perfectly. On the other hand, if there is always harvested energy sufficient to supply the appliances, i.e., $M=N$ and $\mathrm{Pr}\{Z=N\}=1$, then $Y_{i}=0$ for $\forall i$, and we have $I_{p}=0$. When $I_{p}=0$ we say that $\emph{perfect privacy}$ is achieved. Basically, as we harvest more and more energy, we reduce our dependence on the grid energy, and decrease the information leaked to the UP about our real energy consumption. However, note that, at each time instant harvested energy that is not used by the consumer is wasted. For example, when $\mathrm{Pr}\{Z=N\}=1$, we have $E_{w}=N-E[X]$ while $E_{w}=0$ when $\mathrm{Pr}\{Z=0\}=1$. In other words, there is a trade-off between privacy and energy efficiency provided by the EH unit. Introducing an RB into this system will have a dual use and improve this trade-off. RBs can act as a filter for the energy usage profile and decrease $I_{p}$ further while reducing the wasted energy at the same time.

Due to the discrete time nature of the system, it can be represented by a finite state model (FSM)~\cite{RechargeableBattery}. The FSM representation of the system with all the transitions and states evolving as a Markov chain depends on the input load level $N$, the output load level $L$, the harvested energy level $M$ and the RB capacity $K$. As we have mentioned earlier, we consider energy management policies that depend only on the current input load $X_i$, harvested energy $Z_i$, and the previous battery state $B_{i-1}$\footnote{In~\cite{RechargeableBattery} in addition to battery conditioned policies, battery/output load conditioned policies are also studied. However, the authors indicate that they have not found any battery/output load conditioned policy that performs better than the optimal policy that acts solely based on the battery state. We have made the same observation in our numerical analysis.}. We have $s\triangleq(K+1)$ states in our FSM, where state $b_i$ denotes the state of the RB, i.e., the amount of energy stored in the RB at time $i$. We assume $b_{0}=0$. The battery conditioned transitions occur from state $b_{i}$ to $b_{i+1}$ depending on the battery state $b_{i}$, the input load $x_{i+1}$ and the harvested energy $z_{i+1}$. The FSM is simply a Markov chain, and the transitions specify the map to proceed in the chain. Possible transitions are depicted in Fig.~\ref{fig:StateDiagram} for different $(x,z,y)$ triplets and transition probabilities.

\subsection{A Simplified Binary Model} \label{ss:binary_model}

Similarly to~\cite{RechargeableBattery} to keep the presentation and the numerical analysis simple, we initially consider a binary model; that is, we assume $N=L=M=K=1$. However, we note here that the following arguments and evaluation techniques extend to non-binary models directly. From a practical perspective, this binary model corresponds to a system with a single appliance that can be ON or OFF at various time instants with a certain probability, and both the capacity of the RB and the energy generated by the EH are equivalent to the energy used by this device when it is ON. In Sections \ref{s:EffectofBatteryCapacityonPrivacy} and \ref{s:PrivacyatExpenseofWastingGridEnergy} we will consider non-binary battery capacity cases as well.

While the energy management policies can be time-varying in general, we consider time-invariant fixed policies in which the transition probabilities and parameters of the policy are fixed throughout the operation. The probability distributions of the input load and the harvested energy are chosen as Bernoulli distributions, i.e., $\mathrm{Pr}\{X=1\}=p_{x}$ and $\mathrm{Pr}\{Z=1\}=p_{z}$, respectively. The output load $Y^n$ is also a binary sequence which can provide $0$ or $1$ units of energy to the input load at any time instant $i$. Battery state $b_{i}=0$ denotes that the RB is empty while $b_{i}=1$ denotes that the RB is fully charged at time instant $i$. We assume that within each time duration, $i$ to $i+1$, the RB can be charged to battery state, $b_{i}=1$, discharged to battery state, $b_{i}=0$, or remain in the same state depending on the transition probabilities. We do not take into consideration the charging and discharging rates of the RB, and assume that this time duration is enough for fully charging or discharging.

\begin{figure}
\hspace{-0.25cm}
\includegraphics[scale=0.257]{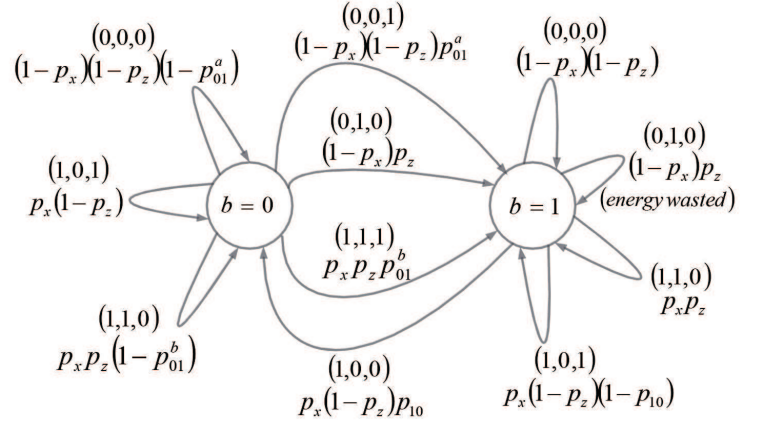}
\caption{Finite state diagram for the battery conditioned energy management policy with $s=2$ states. Each triplet in the figure corresponds to the $(x,z,y)$  values for the corresponding transition. Transition probabilities are also included in the figure.} \label{fig:StateDiagram}
\end{figure}

Let the RB be discharged at time instant $i$, i.e., $b_{i}=0$. There are six possible transitions that can occur as illustrated in Fig.~\ref{fig:StateDiagram}. If the appliances demand zero energy and no energy is harvested, i.e., $(x_{i+1}=0,z_{i+1}=0)$, the EMU chooses either to charge the RB by asking energy from the UP, i.e., $(y_{i+1}=1, b_{i+1}=1)$ with probability $p^{a}_{01}$, or keeps the RB discharged, i.e., $(y_{i+1}=0, b_{i+1}=0)$ with probability $(1-p^{a}_{01})$. If the appliances demand zero energy and one unit of energy is harvested, i.e., $(x_{i+1}=0,z_{i+1}=1)$, the UP does not provide any energy to prevent waste and the RB is charged with harvested energy, i.e., $(y_{i+1}=0, b_{i+1}=1)$. If the appliances demand one unit of energy and no energy is harvested, i.e., $(x_{i+1}=1,z_{i+1}=0)$, the UP must provide one unit of energy to fulfill the energy demand and the RB remains discharged, i.e., $(y_{i+1}=1, b_{i+1}=0)$. If the appliances demand one unit of energy and one unit of energy is harvested at the same time, i.e., $(x_{i+1}=1,z_{i+1}=1)$, either the RB is charged by means of the output load, i.e., $(y_{i+1}=1, b_{i+1}=1)$ with probability $p^{b}_{01}$, or it remains discharged, i.e., $(y_{i+1}=0, b_{i+1}=0)$ with probability $(1-p^{b}_{01})$.

Similarly, let the RB be charged at time instant $i$, i.e., $b_{i}=1$. In this case, there are five possible transitions that can occur as depicted in Fig.~\ref{fig:StateDiagram}. If the appliances demand zero energy and no energy is harvested, i.e., $(x_{i+1}=0,z_{i+1}=0)$, the UP does not provide energy so as not to cause waste and the RB remains charged, i.e., $(y_{i+1}=0, b_{i+1}=1)$. If the appliances demand zero energy and one unit of energy is harvested, i.e., $(x_{i+1}=0,z_{i+1}=1)$, the UP is not expected to provide any energy and the RB remains charged, i.e., $(y_{i+1}=0, b_{i+1}=1)$, while the harvested energy is wasted in this situation. If the appliances demand one unit of energy and no energy is harvested, i.e., $(x_{i+1}=1,z_{i+1}=0)$, the EMU chooses between keeping the RB charged, i.e., $(y_{i+1}=1, b_{i+1}=1)$ with probability $(1-p_{10})$, or discharging it, i.e., $(y_{i+1}=0, b_{i+1}=0)$ with probability $p_{10}$. If the appliances demand one unit of energy and one unit of energy is harvested, i.e., $(x_{i+1}=1,z_{i+1}=1)$, there is no need to ask for energy from the UP and the RB remains charged, i.e., $(y_{i+1}=0, b_{i+1}=1)$.

\section{Information Leakage Rate Computation}\label{s:InformationLeakageRatesec}
In this section we focus on the computation of the information leakage rate, $I_{p}$. From an information theoretic perspective the operation of the EMU which decides on the energy flow in the system using the EH and RB units resembles data compression where the compression is accomplished through a finite state machine. In this analogy, the input load $X^{n}$ corresponds to an i.i.d. data sequence to be compressed, and the output load $Y^{n}$ is the compressed version. The problem is similar to a rate-distortion problem in which the goal is to minimize the mutual information between the source sequence and the compressed version while satisfying the distortion requirement. In our model, the energy provided from the EH device is similar to a distortion requirement. While we want to minimize the mutual information between the original data sequence and the compressed version, we are limited by the allowed distortion, the available harvested energy in our case. A different rate-distortion approach for the SM privacy problem is taken in~\cite{Sankar}. In~\cite{Sankar} the SM is allowed to introduce a certain amount of distortion to its readings before reporting them to the UP, while in our setting distortion is introduced on the real energy consumption values, making the rate-distortion formulation less explicit. See~\cite{PrivacyAlternativeEnergySource} for more on the connection with the rate-distortion theory, where a single-letter information theoretic expression is obtained for the optimal privacy in the absence of an RB. Due to the memory introduced into the system through the battery, a single letter expression is elusive for our problem. However, for a fixed EMU policy, the information leakage rate $I_{p}$ between the input and the output loads can be estimated numerically using the computation method studied in~\cite{InformationRate}. In the following we summarize this computation method.

\begin{figure}
\hspace{-0.27cm}
\includegraphics[scale=0.26]{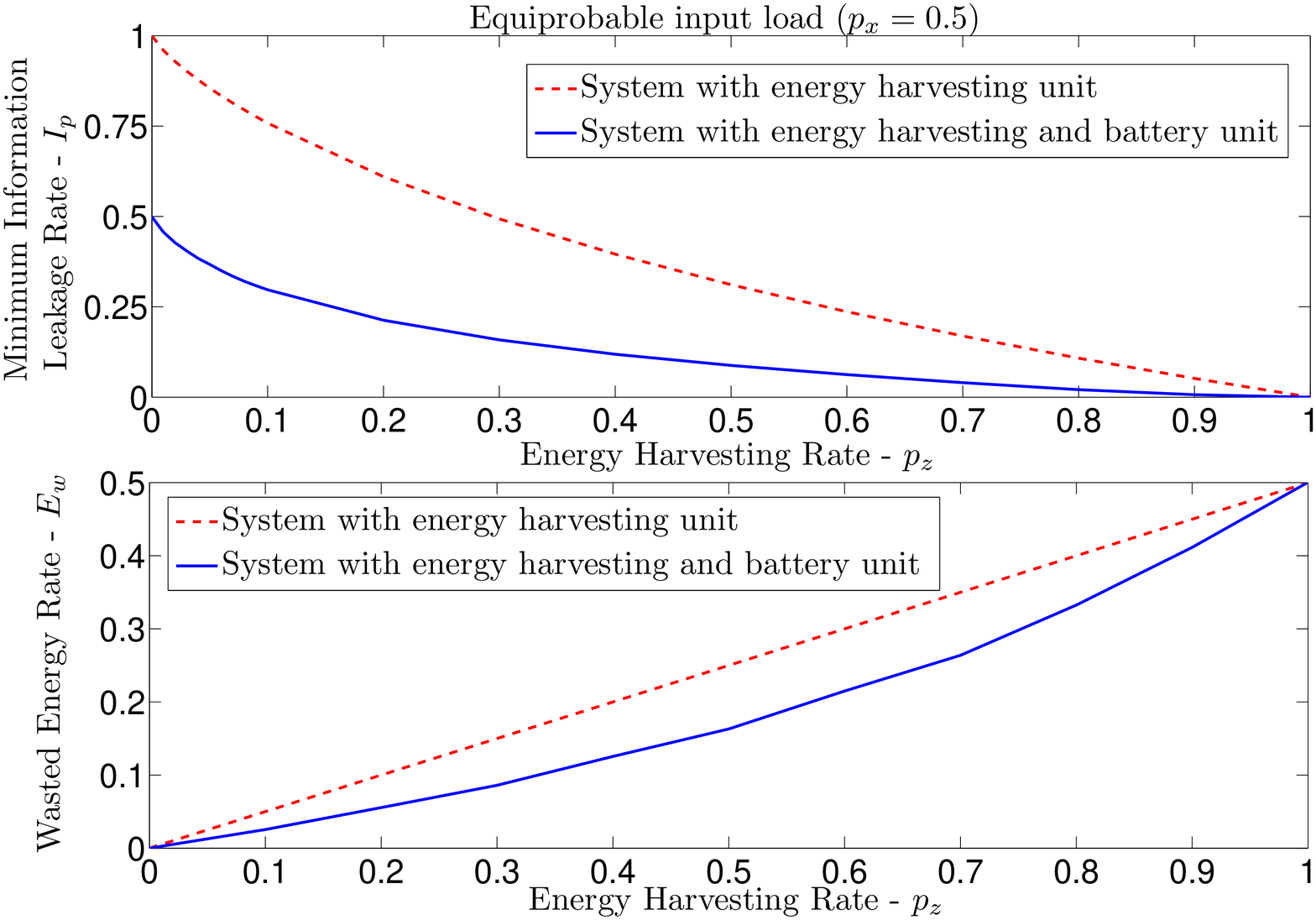}
\caption{Minimum information leakage rate, $I_{p}$, and the corresponding wasted energy rate, $E_{w}$, with respect to harvested energy rate for an EH system with and without an RB.}
\label{fig:WithandWithoutBatteryCompare}
\end{figure}

We first set the values for the transition probabilities and the number of states $s$ in the FSM. For instance, we specify $\{p^{a}_{01}, p^{b}_{01},p_{10}\}$ labeled on Fig.~\ref{fig:StateDiagram} for $s=2$, i.e., $b_{i}\in\left\{{0,1}\right\}$. Afterwards, we sample very long sequences (large $n$) of $X^{n}$, $Z^{n}$ and $Y^{n}$ by using the FSM. We then compute $p(y_{1},y_{2},\cdots,y_{n})$ and $p(x_{1},x_{2},\cdots,x_{n},y_{1},y_{2},\cdots,y_{n})$. Finally, the information leakage rate $I_{p}$ between $X^{n}$ and $Y^{n}$ is estimated as follows:
\begin{align}
 I_{p}&=\frac{1}{n}\big[H(X^n)+H(Y^n)-H(X^n,Y^n)\big]\nonumber \\
&\approx H(X)-\frac{1}{n}\log{p(y_{1},y_{2},\cdots,y_{n})}\nonumber \\
&+\frac{1}{n}\log{p(x_{1},x_{2},\cdots,x_{n},y_{1},y_{2},\cdots,y_{n})}.
\end{align}

The FSM can be represented as a trellis diagram  with the state sequence $\{s_{0},s_{1},\cdots,s_{n}\}$ for the computation of the probabilities $p(y_{1},y_{2},\cdots,y_{n})$ and $p(x_{1},x_{2},\cdots,x_{n},y_{1},y_{2},\cdots,y_{n})$. This computation is basically the forward sum-product recursion of the BCJR algorithm~\cite{BCJR}. We define the state metrics as follows:

\begin{align}
&\mu_{k}(s_{k})\triangleq p(s_{k},y_{1},y_{2},\cdots,y_{k}),\\
&\nu_{k}(s_{k})\triangleq p(s_{k},x_{1},x_{2},\cdots,x_{k},y_{1},y_{2},\cdots,y_{k}).
\end{align}

\noindent Initially, we set the state metrics as follows:

\begin{align}
&\mu_{0}(0)=1,\hspace{0.1 cm}\nu_{0}(0)=1,\hspace{0.1 cm}\mu_{0}(m)=0,\hspace{0.1 cm}\nu_{0}(m)=0,\hspace{0.1 cm}\mbox{for $m\neq0$}. \nonumber
\end{align}

Here, we emphasize that the initial values of the state metrics do not affect the final values of $p(y_{1},y_{2},\cdots,y_{n})$ and $p(x_{1},x_{2},\cdots,x_{n},y_{1},y_{2},\cdots,y_{n})$ due to the convergence for long sequences.

We then compute the state metrics recursively using the transition probabilities $p(x_{k+1},z_{k+1},y_{k+1},s_{k+1}|s_{k})$. For the binary system we use the transition probabilities labeled in Fig.~\ref{fig:StateDiagram}. We have,

\small\begin{align}
&\mu_{k+1}(s_{k+1})=\sum\limits_{z_{k+1}}\sum\limits_{x_{k+1}}\sum\limits_{s_{k}}{\mu_{k}(s_{k})p(x_{k+1},z_{k+1},y_{k+1},s_{k+1}|s_{k})},\\
&\nu_{k+1}(s_{k+1})=\sum\limits_{z_{k+1}}\sum\limits_{s_{k}}{\nu_{k}(s_{k})p(x_{k+1},z_{k+1},y_{k+1},s_{k+1}|s_{k})}.
\end{align}
\normalsize

\noindent We can compute the probabilities $p(y_{1},y_{2},\cdots,y_{n})$ and $p(x_{1},x_{2},\cdots,x_{n},y_{1},y_{2},\cdots,y_{n})$ as the sum of all the final state metrics as follows:
\begin{align}
p(y_{1},y_{2},\cdots,y_{n})&=\sum\limits_{s_{n}}{\mu_{n}(s_{n})},\\
p(x_{1},x_{2},\cdots,x_{n},y_{1},y_{2},\cdots,y_{n})&=\sum\limits_{s_{n}}{\nu_{n}(s_{n})}.
\end{align}

\noindent For large $n$ values, the state metrics $\mu_{k}(\cdot)$ and $\nu_{k}(\cdot)$ tend to zero. Therefore, in practice the recursion is computed with scale factors as follows:

\scriptsize\begin{align}
&\mu_{k+1}(s_{k+1})=\lambda_{\mu_{k+1}}\sum\limits_{z_{k+1}}\sum\limits_{x_{k+1}}\sum\limits_{s_{k}}{\mu_{k}(s_{k})p(x_{k+1},z_{k+1},y_{k+1},s_{k+1}|s_{k})},\\
&\nu_{k+1}(s_{k+1})=\lambda_{\nu_{k+1}}\sum\limits_{z_{k+1}}\sum\limits_{s_{k}}{\nu_{k}(s_{k})p(x_{k+1},z_{k+1},y_{k+1},s_{k+1}|s_{k})},
\end{align}
\normalsize

\noindent where positive scale factors $\{\lambda_{\mu_{1}},\lambda_{\mu_{2}},\cdots,\lambda_{\mu_{n}}\}$ and $\{\lambda_{\nu_{1}},\lambda_{\nu_{2}},\cdots,\lambda_{\nu_{n}}\}$ are chosen such that,

\begin{align}
\sum\limits_{s_{n}}{\mu_{n}(s_{n})}=1,\\
\sum\limits_{s_{n}}{\nu_{n}(s_{n})}=1.
\end{align}

\noindent Finally, the joint probabilities can be computed from the following equations:
\begin{align}
&-\frac{1}{n}\log{p(y_{1},y_{2},\cdots,y_{n})}=\frac{1}{n}\sum\limits_{i=1}^{n}{\log{\lambda_{\mu_{i}}}},\\
&-\frac{1}{n}\log{p(x_{1},x_{2},\cdots,x_{n},y_{1},y_{2},\cdots,y_{n})}=\frac{1}{n}\sum\limits_{i=1}^{n}{\log{\lambda_{\nu_{i}}}}.
\end{align}

We note here that this computation method applies to any discrete model, including an input load with memory, and is not limited to the binary system model considered in this paper. However, identification of the optimal system parameters becomes computationally intractable with an increase in the size of the input and output alphabets, or the battery size.

\begin{figure}
\hspace{-0.6cm}
\includegraphics[scale=0.265]{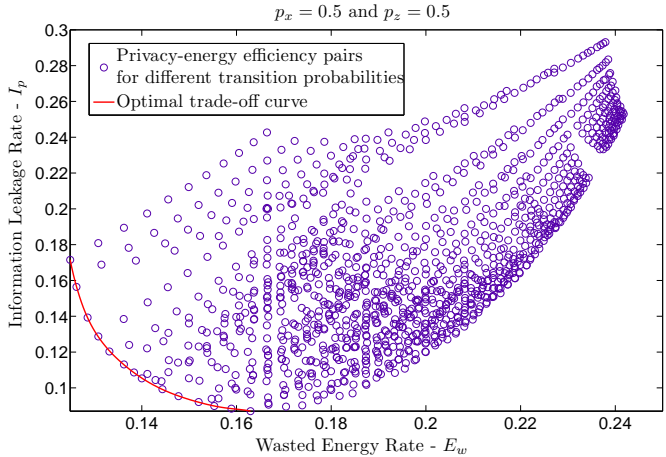}
 \caption{Information leakage rate, $I_{p}$, versus wasted energy rate, $E_{w}$, for $p_{x}=0.5$ and $p_{z}=0.5$.}
\label{fig:MutualandWasted}
\end{figure}

\section{Results and Observations}\label{s:ResultsandObservations}
In this section, we analyze the trade-off between the information leakage rate and energy efficiency numerically using the computation method presented in Section~\ref{s:InformationLeakageRatesec}. Based on these numerical results we provide various observations and conclusions regarding the optimal operation of the EMU from a joint privacy-energy efficiency perspective. In our simulations we focus on the binary model illustrated in Fig.~\ref{fig:StateDiagram}. We focus on a binary system for its simplicity, as otherwise, the transitions in the state diagram get very complicated and the numerical computation outlined in Section~\ref{s:InformationLeakageRatesec} becomes intractable. Later in Section~\ref{s:EffectofBatteryCapacityonPrivacy} we also consider the system with $K>2$ in the absence of an EH unit, and study the effects of the battery capacity on the performance. Furthermore, in Section~\ref{s:PrivacyatExpenseofWastingGridEnergy} we consider a system with high privacy requirements in the absence of an EH unit, and allow the user to waste grid energy in order to increase privacy. In our simulations, we perform an exhaustive search by varying the transition probabilities in Fig.~\ref{fig:StateDiagram} with $0.1$ increments and calculate the information leakage rate for each EMU policy. We use $n=10^6$ for the computations.

\subsection{Effects of energy harvesting rate on privacy and energy efficiency}\label{s:Effectofenergyharvestingrate}
We illustrate the effects of EH rate on both privacy and energy efficiency for an EH system with and without an RB, and also show how privacy and energy efficiency change in the presence of an RB. Fig.~\ref{fig:WithandWithoutBatteryCompare} illustrates the minimum information leakage rate $I_{p}$ and the corresponding wasted energy rate $E_{w}$ with respect to the EH rate $p_{z}$ for an EH system with and without an RB. The results are obtained for an equiprobable input load $p_{x}=0.5$ and different $p_{z}$ values. In a system with an EH device the privacy improves with increasing values of $p_{z}$. This is expected since more energy is provided from the energy harvester as $p_{z}$ increases; and hence, the UP can learn less about the actual energy consumption of the user. On the other hand, an increase in the EH rate leads to an increase in the wasted energy rate as well. This is due to the independence of the energy generation process and the input load. When the EH device harvests a unit of energy, if there is no demand from the appliances and the RB is already charged, this harvested energy will be wasted. Therefore, we can easily notice the trade-off between the information leakage rate $I_{p}$ and the wasted energy rate $E_{w}$ in the system when there is no storage unit.

Comparing the two curves in Fig.~\ref{fig:WithandWithoutBatteryCompare}, we observe that introducing an RB into the system improves the trade-off to a certain extent. It reduces both the minimum information leakage rate $I_{p}$ and the corresponding wasted energy rate $E_w$. When there is no energy harvesting, i.e, $p_{z}=0$, the system reduces to the model studied in~\cite{RechargeableBattery}. In this case, the minimum information leakage rate is found to be $I_{p}=0.5$ for $p_{x}=0.5$. However, when there is an alternative energy source in the system, i.e., $p_z\neq0$, the information leakage rate can be reduced significantly. The EH rate can be considered as a system parameter that defines the achievable privacy-energy efficiency trade-off, and needs to be chosen by the system designer depending on the input load and the desired operating point.

\begin{figure}
\hspace{-0.6cm}
\includegraphics[scale=0.265]{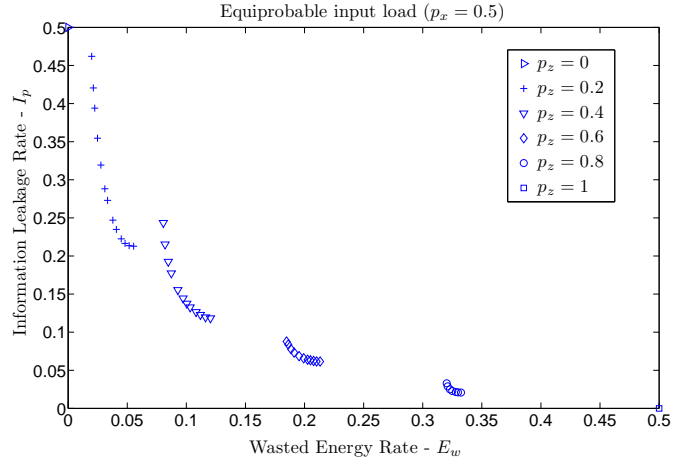}
\caption{The Pareto optimal $\big(I_{p},E_{w}\big)$ pairs for $p_{x}=0.5$ and for different $p_{z}$ values. Optimal pairs for different $p_{z}$ values are illustrated with different markers.}
\label{fig:MutualWastedForDifferentpz}
\end{figure}

\subsection{Privacy-energy efficiency trade-off}\label{s:Privacy-Energy efficiency trade-off}

\begin{table}[ht]
\begin{center}
\caption{RESULTS FROM THE TRADE-OFF PAIRS FOR DIFFERENT $p_{z}$ VALUES} 
\begin{tabular}{c|c c|c c} 
$p_{z}$ & $\min I_{p}$ & $E_{w}$ for $\min I_{p}$ & $\min E_{w}$ & $I_{p}$ for $\min E_{w}$ \\
[0.5ex] 
\hline\hline 
0 & 0.5 & 0 & 0 & 0.5  \\ [0.1ex]   
0.2 & 0.213 & 0.055 & 0.02 & 0.462  \\ [0.1ex]
0.4 & 0.118 & 0.12  & 0.081 & 0.243  \\ [0.1ex]
0.6 & 0.062 & 0.213 & 0.185 & 0.088 \\ [0.1ex]
0.8 & 0.02 & 0.332 & 0.32 & 0.032 \\ [0.1ex]
1 & 0 & 0.5 & 0.5 & 0 \\ 
\hline\hline 
\end{tabular}
\label{table:table1} 
\end{center}
\end{table}

In Section~\ref{s:Effectofenergyharvestingrate} we have found the wasted energy rate corresponding to the battery policy that minimizes the information leakage rate. Here, we characterize the whole trade-off between the privacy and energy efficiency for given EH rates. The trade-off for the values of $p_{x}=p_{z}=0.5$ is illustrated in Fig.~\ref{fig:MutualandWasted}. Each circle in the figure marks an $\big(I_{p},E_{w}\big)$ pair that can be achieved by assigning different transition probabilities labeled on Fig.~\ref{fig:StateDiagram}. The Pareto optimal trade-off curve is the one that is formed by the points on the lower-left corner of the figure, i.e., the points for which  $I_{p}$ and $E_{w}$ cannot be improved simultaneously. The minimum information leakage rate value is $I_{p}=0.088$ for which we have $E_{w}=0.163$. The minimum wasted energy rate is $E_{w}=0.125$ for which we have $I_{p}=0.171$. These two pairs correspond to the corner points of the trade-off curve in Fig.~\ref{fig:MutualandWasted}. According to the requirements of the system, the operating point can be chosen anywhere on the trade-off curve. Note that, we can apply a convexification operation on the set of achievable $(I_p, E_w)$ pairs using time-sharing arguments.

We also study the trade-off between the information leakage rate, $I_{p}$, and the wasted energy rate, $E_{w}$, for different $p_{z}$ values to observe the effect of the EH rate on the achievable privacy-energy efficiency trade-off. Fig.~\ref{fig:MutualWastedForDifferentpz} illustrates the Pareto optimal $\big(I_{p},E_{w}\big)$ pairs for $p_{x}=0.5$ and for different $p_{z}$ values. Each marker in the figure marks an $\big(I_{p},E_{w}\big)$ pair achieved by assigning different transition probabilities, and we include only the points that are not Pareto dominated by any other point. We obtain a different privacy-energy efficiency trade-off for each $p_{z}$ value as illustrated in Fig.~\ref{fig:MutualWastedForDifferentpz}. The corner points of these trade-off curves are listed in Table~\ref{table:table1} for different $p_{z}$ values. Since there is no harvested energy in the system for $p_{z}=0$, there is no wasted energy and as a result, the optimal operating point is found as the minimum information leakage rate, $I_{p}=0.5$ and wasted energy rate, $E_{w}=0$, which is the same as the model studied in~\cite{RechargeableBattery}. Note that while the minimum information leakage rate decreases with increasing values of $p_{z}$, the minimum wasted energy rate increases. When energy is harvested with $p_{z}=1$, the optimal point is found to be $I_{p}=0$ and $E_{w}=0.5$, that is, perfect privacy can be achieved at the expense of wasting half of the harvested energy on average. In this case, there is no information leakage since the user never asks energy from the UP and the wasted energy rate converges to $Pr\{X=0\}=1-p_{x}$.

\begin{figure}
\centering
\includegraphics[scale=0.23]{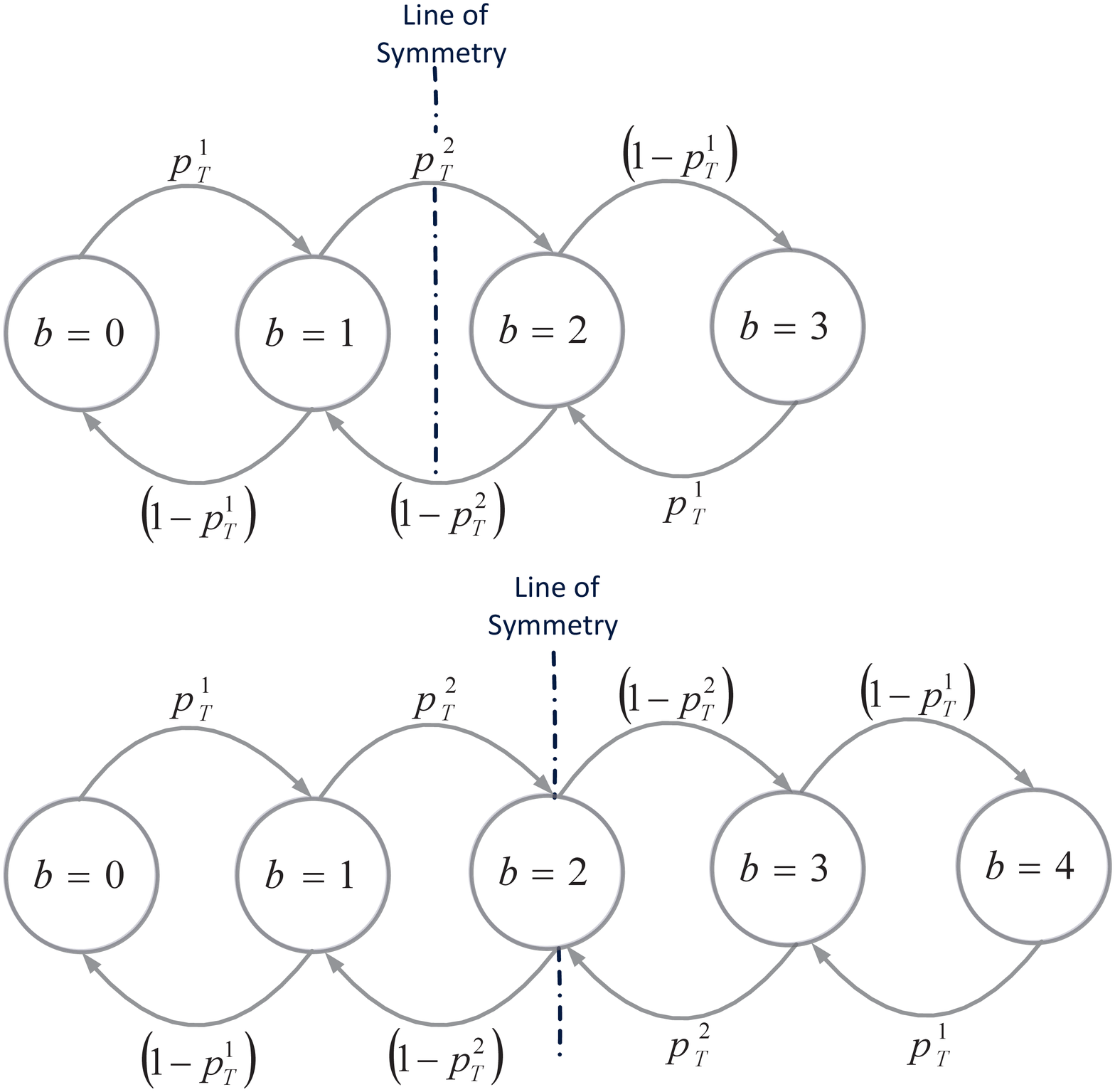}
\caption{Finite state diagrams for battery-conditioned energy management policies with battery capacities $K=3$ and $K=4$. Symmetric and complementary transition probabilities are illustrated for the computation of the minimum information leakage rate in case of an equiprobable input load, i.e., $p_{x}=0.5$.}
\label{fig:StateDiagramSymmetry}
\end{figure}

We also study biased input loads by considering the two cases with $p_{x}=0.89$ and $p_{x}=0.11$, which we call the $\emph{heavy load}$ and $\emph{light load}$ scenarios. The entropy rate of the input load for both the heavy and light load cases is $H(X)=0.5$. Note that the input load is biased towards $X=1$ for the heavy load system, i.e., the appliances are more likely to demand energy. For the heavy load case when we do not have an EH unit in the system, i.e., $p_{z}=0$, we find the minimum information leakage rate to be $I_{p}=0.23$~\cite{RechargeableBattery}. When there is an energy harvester in the system with $p_{z}=0.5$, the minimum information leakage rate reduces significantly to $I_{p}=0.026$ while the corresponding wasted energy rate is $E_{w}=0.043$. The minimum wasted energy rate is obtained as $E_{w}=0.011$ for which we have $I_{p}=0.105$. It is obvious that wasting energy is less likely in the heavy load case. The energy is wasted only when we have ${b_{i}=1,x_{i+1}=0,z_{i+1}=1}$ as shown in Fig.~\ref{fig:StateDiagram}. Thus, when the appliances have higher energy demands, the user is less likely to face the condition for energy wasting. Similarly, in the light load case, i.e., $p_{x}=0.11$, $E_{w}$ increases as less energy is required by the appliances. For example, the minimum information leakage rate is found to be $I_{p}=0.027$ with $E_{w}=0.088$, and the minimum wasted energy rate is found to be $E_{w}=0.087$ for $I_{p}=0.03$. We observe that both the heavy and light load systems can achieve almost the same level of maximum privacy while the wasted energy rate of the light load system is double the rate of the heavy load system at this point of operation.

\subsection{Effects of battery capacity on privacy}\label{s:EffectofBatteryCapacityonPrivacy}
\begin{figure}
\hspace{-0.5cm}
\includegraphics[scale=0.265]{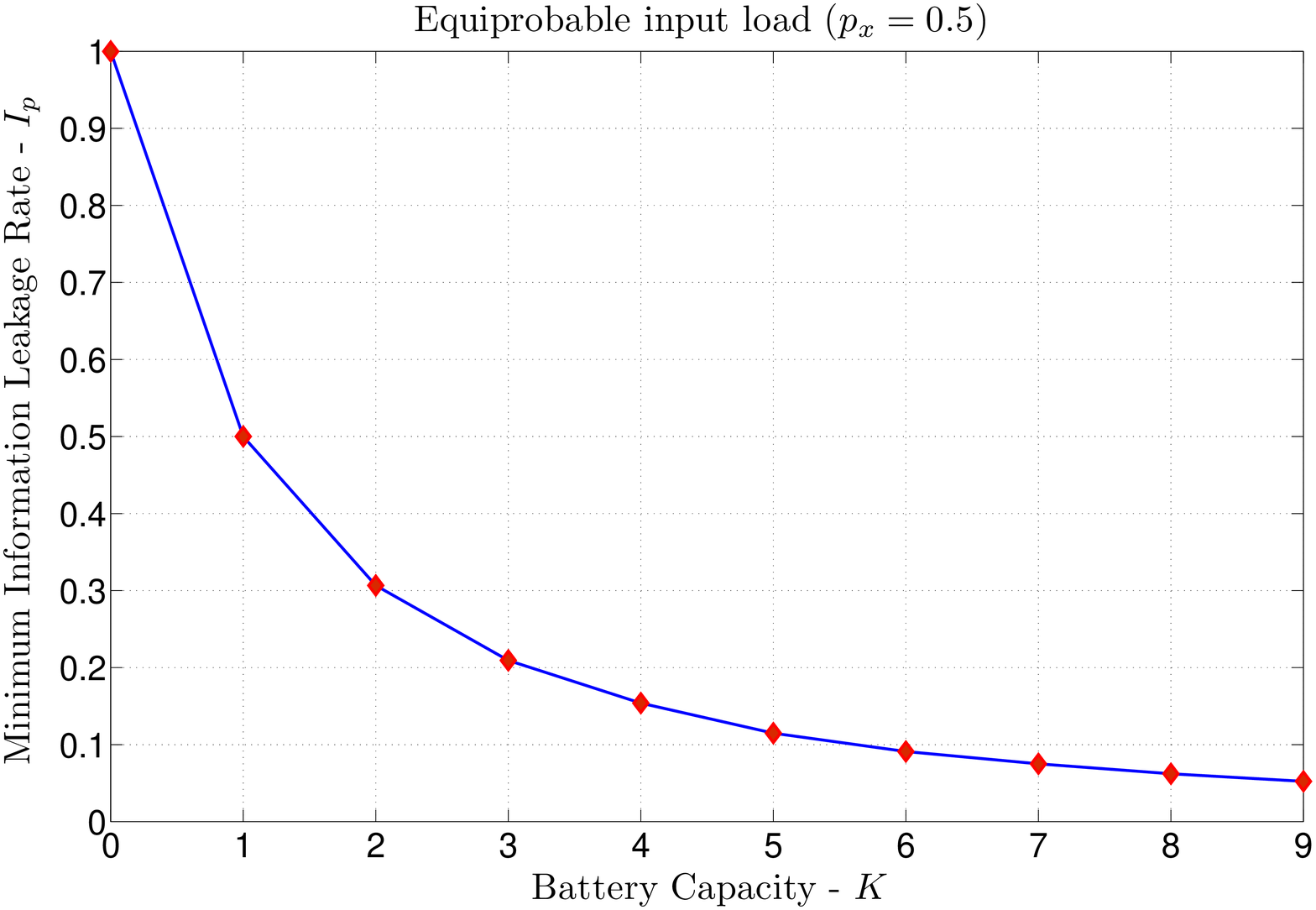}
\caption{Minimum information leakage rate, $I_{p}$, versus battery capacity, $K$.}
\label{fig:MutualvsBattery}
\end{figure}

We have observed that alternative energy sources can help reduce the information leakage rate significantly while RBs help improve the energy efficiency as well as privacy. Next, we study the effects of the RB capacity on privacy. It is expected that if we increase the RB capacity $K$, the trade-off curve illustrated in Fig.~\ref{fig:MutualandWasted} will move toward the origin, i.e., the privacy and energy efficiency will be improved simultaneously. For example, in the asymptotic limit of infinite storage capacity, perfect privacy can be achieved by charging the battery initially, and never asking for any energy from the UP afterwards. To highlight the effects of the battery capacity on the achievable privacy we consider an RB with capacity $K$, and no EH device. While the complexity of the numerical analysis grows quickly with the battery size, we have observed that for an equiprobable input load, i.e., $p_{x}=0.5$, there is a symmetry and complementarity among the optimal transition probabilities in the finite state diagram which significantly reduces the computation time of the minimum information leakage rate. The minimum information leakage rate is achieved when, $1)$ the sum of transition probabilities between two states is equal to one, and $2)$ there is a symmetry in the transition probabilities of the two sides of the finite state diagram separated by the line of symmetry. Fig.~\ref{fig:StateDiagramSymmetry} depicts this symmetry and complementarity on a finite state diagram for battery capacity $K=3$ and $K=4$, respectively. Using this observation which reduces the complexity of the computation, we have increased the battery capacity $K$ and obtained the minimum information leakage rates corresponding to different values of $K$. For moderate battery capacity values Fig.~\ref{fig:MutualvsBattery} illustrates the effects of the battery capacity on the minimum information leakage rate $I_{p}$ for $p_{x}=0.5$. The minimum information leakage rate falls below $0.1$ even with an RB of $6$ units of capacity. This result shows that even a small increase in the RB capacity leads to a significant reduction in the minimum information leakage rate. As RB capacity increases more, the minimum information leakage rate $I_{p}$ continues to decrease, but with a decreasing slope.

\subsection{Privacy at the expense of wasting grid energy}\label{s:PrivacyatExpenseofWastingGridEnergy}
We have already shown that whenever the user has higher privacy requirements, the system with EH and RB units can provide strong privacy assurances by simply increasing the EH rate, $p_{z}$. When there is no EH unit in the system, we need to increase the capacity of the RB to cope with high privacy requirements. However, increasing the capacity of the RB can be costly or even physically impossible. In this case the privacy of the user can be improved by allowing the user to demand energy from the UP even when there is no energy demand from the appliances, i.e., $x_{i}=0$, and the RB is already full, i.e., $b_{i}=K$. Through wasting additional energy from the UP, which is likely to be more expensive than the harvested energy, the energy consumption profile of the appliances can be further hidden from the UP and privacy can be increased up to perfect privacy by increasing the energy waste level.

\begin{figure}
\hspace{-0.5cm}
\includegraphics[scale=0.265]{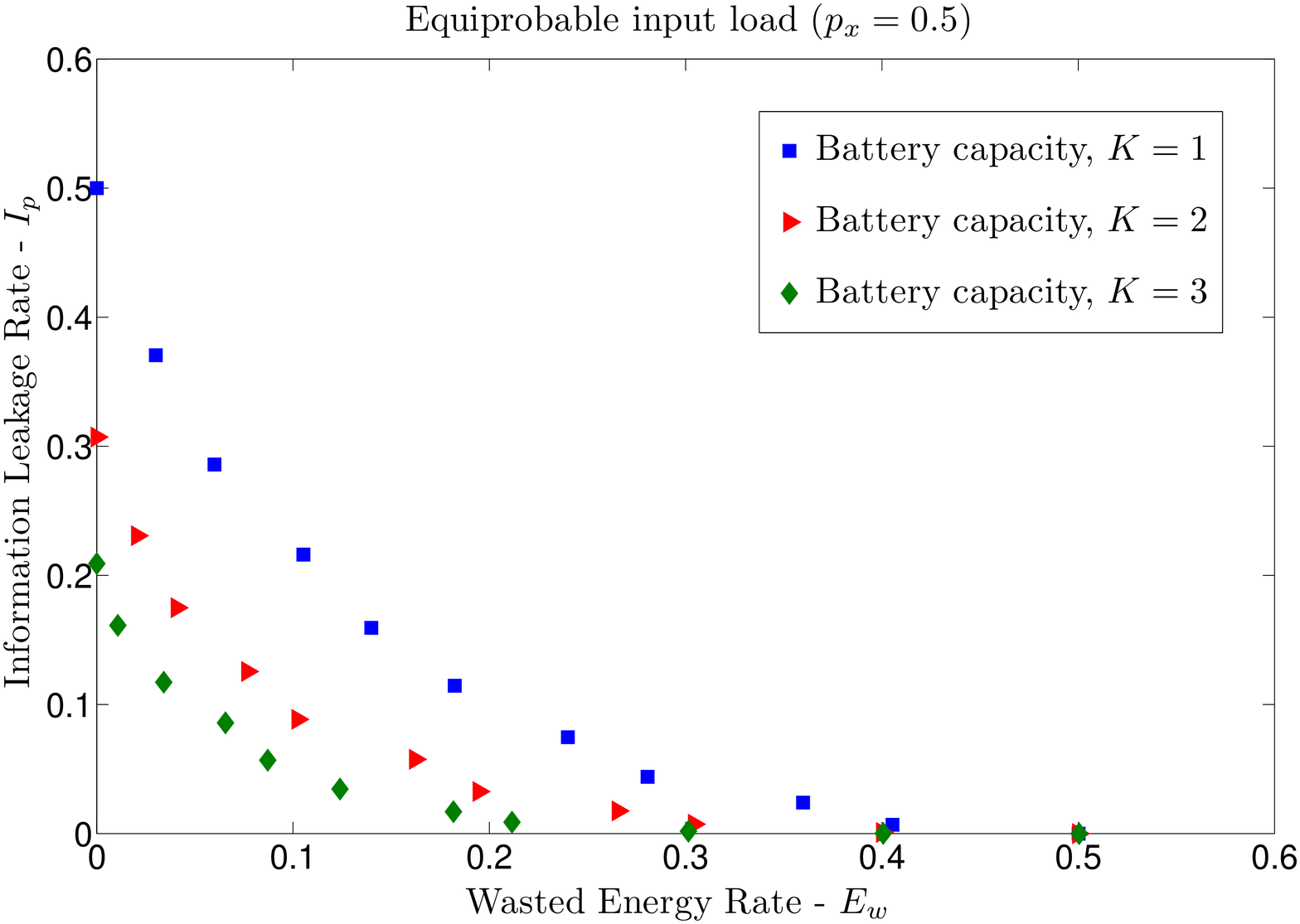}
 \caption{Information leakage rate, $I_{p}$,  versus wasted energy rate, $E_{w}$, for the case of wasting grid energy.}
\label{fig:MutualandWastedAllowed}
\end{figure}

To study the effects of wasting grid energy on privacy, we consider battery conditioned policies with binary input/output load values and an RB with capacity of $K$ units. Let RB be fully charged at time instant $i$, i.e., $b_{i}=K$. Even if the appliances do not consume any energy at time instant $i+1$, i.e., $x_{i+1}=0$, we allow the EMU to demand energy from the UP, i.e., $y_{i+1}=1$, with probability $p_{w}$, and $y_{i+1}=0$ with probability $(1-p_{w})$. In other words, we allow wasting the grid energy with probability $p_{w}$, by which we obscure the information of the UP about the real energy consumption. Fig.~\ref{fig:MutualandWastedAllowed} illustrates the achievable points on the $\big(I_{p},E_{w}\big)$ trade-off, obtained for an equiprobable input load, $p_{x}=0.5$, and for increasing RB capacity values, $K=1$, $K=2$, and $K=3$. In this simulation, to keep the simulation time reasonable we find the achievable points for each capacity value $K$, by considering only complementary transition probabilities as depicted in Fig.~\ref{fig:StateDiagramSymmetry}, such that the sum of the transition probabilities between two states is equal to $1$. Moreover, we compute the wasted energy rate by using Eqn.~(\ref{eq:wastedenergyrate}), but we choose $Z_{i}=0$ in the equation since there is no EH unit in the current scenario. We can see that the privacy can be significantly improved by wasting more energy, i.e., by increasing $p_{w}$. For instance, when perfect privacy is required by the system, the information leakage rate can be reduced to zero by wasting energy with $p_{w}=1$. The wasted energy rate converges to $Pr\{X=0\}=1-p_{x}$ on average for $p_{w}=1$, i.e., $E_{w}=0.5$, because we waste energy only when the RB is fully charged, $b_{i}=K$, and there is no input load, $X_{i}=0$. If we increase the RB capacity $K$, as we can see in Fig.~\ref{fig:MutualandWastedAllowed}, both the information leakage rate and the wasted energy rate are improved for the same energy waste probability, $p_{w}$. The operating point on the trade-off curve can be chosen according to the privacy requirement of the system and the cost of energy provided by the UP.

\section{Conclusions}\label{s:ConclusionsFutureWork}
We have studied the privacy-energy efficiency trade-off in smart meter systems in the presence of energy harvesting and storage units. We have considered an EH unit that provides energy packets at each time instant in an i.i.d. fashion, and a finite capacity rechargeable battery that provides both energy efficiency by storing extra energy for future use, and increased privacy by hiding the load signature of the appliances from the utility provider. We have used a finite state model to represent the whole system, and studied the information leakage rate between the input and output loads to measure the privacy of the user from an information theoretic perspective.

We have used a numerical method to calculate the information leakage rate. Due to the memory introduced by the RB, obtaining a closed-form expression for the information leakage rate is elusive. For the sake of simplicity, we have considered binary input and output loads and focused on battery-dependent energy management policies in our simulations, and numerically searched for the energy management strategy that achieves the best trade-off between privacy and energy-efficiency. We have shown that the information leakage rate can be significantly reduced when both an energy harvester and an RB are present. As the EH rate increases, we have observed that the privacy of the system significantly improves. On the other hand, this also increases the amount of wasted energy. For a fixed EH rate, we have numerically obtained the optimal trade-off curve between the achievable information leakage and wasted energy rates. Different points on this trade-off curve can be achieved by changing the stochastic battery policy used by the energy management unit. According to the needs and priorities of the system, an operating point can be chosen on this trade-off curve. We have also obtained the corresponding trade-off curves for different EH rates.

We have studied the effects of the battery capacity on the achievable privacy by focusing on a system with only an RB. We have observed that increasing the capacity of the RB has a significant impact on the reduction of the information leakage rate, and thereby, on the privacy. Moreover, we have examined the wasting of grid energy to fulfill the increased privacy requirements of the user when there is only an RB in the system. We have observed that even in the absence of an EH device and with a finite capacity RB, the privacy level can be increased up to perfect privacy by wasting more energy from the grid.

\bibliographystyle{IEEEtran}
\bibliography{JSAC_FinalArxiv}

\end{document}